\documentclass[]{spie}  

\newcommand*\aap{A\&A}

\newcommand*\aj{AJ}

\newcommand*\apj{ApJ}
\newcommand*\apjl{ApJ}

\newcommand*\apjs{ApJS}

\newcommand*\icarus{Icarus}

\newcommand*\mnras{MNRAS}

\newcommand*\pasp{PASP}

\newcommand*\planss{Planet.~Space~Sci.}

\newcommand*\procspie{Proc.~SPIE}

\usepackage{amsmath,amsfonts,amssymb}
\usepackage{graphicx}
\usepackage{float}
\usepackage[colorlinks=true, allcolors=blue]{hyperref}

\title{Development of the SPECULOOS exoplanet search project}

\author[a]{D. Sebastian}            
\author[b]{P. P. Pedersen}          
\author[b]{C. A. Murray}            
\author[c]{E. Ducrot}               
\author[c]{L. J. Garcia}            
\author[d]{A. Burdanov}             
\author[c,k]{F. J. Pozuelos}        
\author[c,f,k]{L. Delrez}              
\author[e]{R. Wells}
\author[a]{G. Dransfield}
\author[c]{M. Gillon}
\author[e]{B.-O. Demory}
\author[b]{D. Queloz}
\author[a]{A. H.M.J. Triaud}
\author[d]{J. de Wit}
\author[k]{E. Jehin}
\author[g]{Y. G\'omez Maqueo Chew}
\author[i]{M. N. G\"unther}
\author[d]{P. Niraula} 
\author[d]{B. V. Rackham}
\author[e]{N. Schanche}
\author[k]{S. Sohy}
\author[b]{S. Thompson}

\affil[a]{School of Physics \& Astronomy, University of Birmingham, Edgbaston, Birmingham B15 2TT, United Kingdom}

\affil[b]{Cavendish Laboratory, JJ Thomson Avenue, Cambridge CB3 0HE, UK}

\affil[c]{Astrobiology Research Unit, University of Li\`ege, All\'ee du 6 ao\^ut, 19, 4000 Li\`ege (Sart-Tilman), Belgium}

\affil[d]{Department of Earth, Atmospheric and Planetary Sciences, MIT, 77 Massachusetts Avenue, Cambridge, MA 02139, USA}

\affil[e]{University of Bern, Center for Space and Habitability, Gesellschaftsstrasse 6, 3012 Bern, Switzerland}

\affil[f]{Observatoire de l’Université de Genève, Chemin des Maillettes 51, Versoix, CH-1290, Switzerland}

\affil[g]{Instituto de Astronomía, Universidad Nacional Autónoma de Méx- ico, Ciudad Universitaria, Ciudad de México, 04510, México}

\affil[i]{Department of Physics, and Kavli Institute for Astrophysics and Space Research, MIT, Cambridge, MA 02139, USA}

\affil[k]{Space sciences, Technologies and Astrophysics Research (STAR) Institute, Universit{\'e} de Li{\`e}ge, All{\'e}e du 6 Ao{\^u}t 19C, 4000 Li{\`e}ge, Belgium} 

\authorinfo{Further author information:\\ 
Corresponding Author: D.Sebastian, E-mail: D.Sebastian.1@bham.ac.uk\\
SPECULOOS PI: Michael Gillon, E-mail: michael.gillon@uliege.be}

\begin{document} 
\maketitle

\begin{abstract}
SPECULOOS (Search for habitable Planets EClipsing ULtra-cOOl Stars) aims to perform a transit search on the nearest ($<40$\,pc) ultracool ($<3000$K) dwarf stars. The project's main motivation is to discover potentially habitable planets well-suited for detailed atmospheric characterisation with upcoming giant telescopes, like the James Webb Space Telescope (JWST) and European Large Telescope (ELT). The project is based on a network of 1\,m robotic telescopes, namely the four ones of the SPECULOOS-Southern Observatory (SSO) in Cerro Paranal, Chile, one telescope of the SPECULOOS-Northern Observatory (SNO) in Tenerife, and the SAINT-Ex telescope in San Pedro M\'artir, Mexico. The prototype survey of the SPECULOOS project on the 60~cm TRAPPIST telescope (Chile) discovered the TRAPPIST-1 system, composed of seven temperate Earth-sized planets orbiting a nearby (12~pc) Jupiter-sized star. In this paper, we review the current status of SPECULOOS, its first results, the plans for its development, and its connection to the Transiting Exoplanet Survey Satellite (TESS) and JWST.  
\end{abstract}


\section{INTRODUCTION}
\label{sec:intro}  



The past decade of atmospheric studies on close-in giant exoplanets has enabled the first constraints on planetary exospheres, molecular and atomic species, elemental ratios, temperature profiles, clouds, and even atmospheric circulations\cite{Madhusudhan19}. The main laboratories driving this revolution are thousands of transiting exoplanets that have been detected by ground and space-based transit search surveys. The field has developed to perform similar atmospheric studies for smaller and more temperate planets. Examples are GJ 1214b\cite{cha09} (M4.5\,V) or more recently K2-18 b\cite{tsiaras2019,benneke19} (M2.5\,V). For the latter, a temperate `mini-Neptune', the detection of atmospheric water vapour marks an important step towards constraining the atmospheric composition of potentially habitable worlds.
Nevertheless, for temperate, rocky exoplanets that transit solar-type stars, such atmospheric characterisations are far beyond reach with current and upcoming instrumentation. 

This is not true for the latest-type dwarf stars. First, the signal-to-noise ratio (S/N) for eclipse spectroscopy measurements depends on the size-ratio between a planet and its host star and therefore increases for late-type dwarf stars. Second, planets with shorter orbital periods receive the same stellar irradiation as Earth due to the lower luminosities of their host star, resulting in more frequent planetary transits for temperate, rocky planets. 

Studies\cite{Lisa2009,dewit2013,Morley2017} have shown that the combination of both effects brings the atmospheric characterisation of temperate Earth-sized exoplanets within reach of the upcoming James Webb Space Telescope (JWST). This applies under the condition that they transit stars that are both very nearby ($<~ 15$\,pc) and very small ($<~ 0.15$\,$R_{\odot}$), with spectral type M6 or later.

The first detection of terrestrial planets orbiting a M-dwarf was Kepler-42\cite{Muirhead12} (spectral type M4), with its compact system of sub-Earth-sized planets. Other transiting rocky planets were detected around mid-to-late-type M-dwarfs by MEarth\cite{Nutzman2008} (GJ1132\cite{BertaT15} (M4\,V); LHS 1140\cite{dittmann17,ment2019} (M4.5\,V)) and by TESS\cite{Ricker2015} (LP 791-18\cite{crossfield19} (M6\,V); LHS 3844\cite{Vanderspek2019} (M6\,V); TOI 540\cite{Ment2020} (M4\,V)). 
Although, short-period, rocky planets are more common around low-mass stars\cite{howard12,Ullman19} and show deeper and more frequent transits than for earlier-type host-stars, not many of such transiting planets have actually been detected around late-type M-dwarfs. This is especially true for ultracool dwarfs. The classical definition of ultracool dwarfs (hereafter, UCDs) includes dwarfs with spectral type M7 and later, as well as brown dwarfs (BDs) \cite{kirkpatrick97,Kirkpatrick05}. 

The main reason for this apparent lack of planets is the faintness of these stars. Transit search surveys using telescopes with relatively small apertures ($< 50$\,cm), like MEarth and TESS, have a high detection potential for mid-type M-dwarfs (M3 to M5), but this potential drops sharply for later-type objects \cite{Sullivan2015,barclay18}.
The unique system of seven transiting planets that has been detected orbiting TRAPPIST-1\cite{Gillon2016,Gillon2017,Luger2017a}, a nearby (12\,pc) M8V star, is up-to-date the most optimal, laboratory for atmospheric characterisation with JWST in the temperate Earth-sized regime\cite{LustigYaeger2019,macdonald19}.

Due to the scarcity of such detections, not much is known about the structure of planetary systems of late-type M-dwarfs. Until now, TRAPPIST-1 has remained the only transiting system discovered around an UCD, while some statistical constraints could be inferred from the null results of several recent projects \cite{Demory2016,he17,segear19,lienhard20}. 


What kinds of planetary system could be formed around UCDs have been analysed by several theoretical studies\cite{payne07,raymond07,lissauer07,montgomery09,alibert17,coleman19,schoonenberg19,miguel20,Liu20}, which spanned a variety of outcomes from water-rich to water-poor planets, along with several orbital architectures. Notably is the inferred period distribution of planets for typical UCDs. While UCDs are commonly expected to have one or more planets with a period distribution that peaks for short-period planets\cite{coleman19} also a bimodal period distribution seems to be likely\cite{miguel20} with one group with very short periods and another group with larger periods, located beyond the snow line.
A transit search survey, only focusing on short-period planets of UCDs can already distinguish between these various outcomes.

A photometric survey specifically designed to search for transits of Earth-sized planets, or smaller, among the nearest UCDs will allow first, the detection of Earth-like planets amenable for atmospheric characterisation, and second, to draw constraints that could be used to differentiate between different models of planet formation among the lowest-mass stars. 
Given the relative faintness of those objects and the lessons learned from the TRAPPIST-UCDTS prototype survey\cite{Gillon2013,Burdanov2018,lienhard20}, we 
developed SPECULOOS (Search for habitable Planets EClipsing ULtra-cOOl Stars). The SPECULOOS project has been initiated and led by the University of Li\`ege and is done in collaboration with the Universities of Cambridge, Birmingham, Bern, and Massachusetts Institute of Technology. It is based on a network of 1\,m class, robotic telescopes that have a significant detection efficiency for planets as small as the Earth -- and even smaller -- for a large fraction of UCDs within 40\,pc.
The goals of SPECULOOS are: (i) the search for rocky planets that are well-suited for atmospheric characterisation with future facilities like JWST and (ii) more globally, to perform a volume-limited ($<40$\,pc) transit search of UCDs to derive a statistical census of their short-period planet population\cite{Delrez2018b,Gillon2018}.

In this paper, we detail the current status of this survey. In section \ref{sec:network} we describe the observatories that form the ground-based SPECULOOS network, and their remote operations. Section \ref{sec:targets} details the target selection, observing programs and scheduling. In Section \ref{sec:pipeline} we review the data reduction, photometric quality control and impact of atmospheric effects on near-infrared (NIR) observations of UCDs. In Section \ref{sec:portal} we detail the data archive and user interface for the SPECULOOS survey. In section \ref{sec:move} we give an outlook towards a new pipeline, which uses the SPECULOOS main survey to search for moving objects, and finally discuss first scientific results, obtained with the SPECULOOS network in section \ref{sec:discussion}.

\section{Network of ground-based observatories}
\label{sec:network}

The SPECULOOS network (see Table \ref{tab:Multimedia-Specifications}) is composed of six telescopes at three different sites. The SPECULOOS Southern Observatory\cite{Delrez2018b,Jehin2018,murray20} (SSO) with four telescopes named after the four Galilean moons Io, Europa, Ganymede, and Callisto at ESO Paranal Observatory (Chile) became fully operational in January 2019. 
The SPECULOOS Northern Observatory\cite{Niraula2020} (SNO) is currently composed of one telescope (Artemis), which is located at the Teide Observatory (Canary Islands, Spain) and operational since June 2019.
SAINT-EX\cite{Demory2020} (Search And characterIsatioN of Transiting EXoplanets) with one telescope in San Pedro M\'{a}rtir observatory (Mexico) became operational in March 2019.
Each of these observatories is devoting 70\% of its usable observational time to the SPECULOOS survey. The remaining time is used for different programmes. For example, a fraction of this time of the Artemis telescope is dedicated to educational and outreach programs. These programs are carried out with collaborators in the Canary Islands and from the Boston area. A large part of this observing time on SPECULOOS telescopes is devoted to joint annex programmes (see section \ref{sec:discussion}). In addition, the two 60\,cm TRAPPIST\cite{Gillon11,jehin11} robotic telescopes (one in Chile, the other in Morocco), while not officially part of the SPECULOOS network, have devoted a fraction of their time to supporting the project and focus on its brightest targets.

\begin{table}[ht]
\caption{Observatories of the SPECULOOS network of telescopes} 
\label{tab:Multimedia-Specifications}
\begin{center}       
\begin{tabular}{|l|l|l|l|l|l|}
\hline
\rule[-1ex]{0pt}{3.5ex}  Observatory & Telescopes & Host observatory & Coordinates & Height  \\
\hline
\rule[-1ex]{0pt}{3.5ex}  SSO & 4 & ESO Paranal Observatory, Chile  &24.61596 S 70.39057 W& 2490\,m \\
\hline
\rule[-1ex]{0pt}{3.5ex}  SNO & 1 & Teide Observatory, Tenerife, Spain  &28.30000 N 16.51158 W& 2438\,m\\
\hline
\rule[-1ex]{0pt}{3.5ex}  SAINT-EX & 1 & Observatorio Astronómico Nacional, Mexico &31.04342 N 115.45476 W & 2780\,m   \\
\hline
\end{tabular}
\end{center}
\end{table}
Each of the SPECULOOS observatories is composed of identical robotic Ritchey-Chrétien telescopes with 1\,m aperture, designed and built by the German company ASTELCO\footnote{\url{http://www.astelco.com}}. The optical design is comprised of a 1\,m primary mirror with f/2.3 focal ratio and is coupled with a 28\,cm diameter secondary. The combined system focal ratio is f/8. 
Each telescope is equipped with a robotic equatorial ASTELCO New Technology Mount NTM-1000. This mount uses direct-drive torque motors, which allow fast slewing (up to 20 degrees per s), accurate pointing (better than 3 arcseconds) and tracking accuracy better than 2 arcseconds over 15 minutes without any guiding.
It is designed for continuous tracking without meridian flip.
The telescope is installed in a 6.25\,m dome designed and built by the Gambato\footnote{\url{http://www.gambato.com}} company, which is fully integrated to the robotic telescope control system.

Each telescope is equipped with an Andor iKon-L thermoelectrically cooled camera with a near-infrared optimised, deep depletion 2k x 2k e2v CCD detector (13.5 $\mu$m pixel size). The field of view on the sky is 12 x 12 arcminutes, yielding a pixel scale of 0.35~arcseconds/pixel. Exposure control is realised by a mechanical shutter, using overlapping iris blades. Despite the fact that these shutters are generally very durable, they have a limited lifetime. To increase this shutter lifetime, we restrict our remote observations to exposure times larger than 10 seconds. The camera is usually operated at – 60$^{\circ}$ C (via five-stage Peltier cooling) with a dark current of 0.3 electrons/s/pixel. The detector provides high sensitivity in a wide range (350-950 nm), with a maximum quantum efficiency of 94\% at both 420 and 740nm.
Each camera has its own filter wheel providing the Sloan g’, r’, i’, z', and two special exoplanet filters; the near-infrared luminance \textit{I}+\textit{z} filter (transmittance $>$ 90\% from 750 to beyond 1000nm, which is mostly used for the SPECULOOS core program); and a blue-blocking filter called Exo (transmittance $>$ 90\% from 500 to beyond 1000nm).

Every aspect of the remote observations is controlled via ACP Expert Observatory Control Software\footnote{\url{https://acpx.dc3.com}}. It communicates to the camera, the telescope control system via ASCOM\footnote{\url{https://ascom-standards.org}} compatible interface, and provides an inbuilt weather server that allows to continuously monitor data from the weather station. The human intervention is reduced to a quick status check and manual starting of the nights observing plans which is carried out remotely via a secure Virtual Private Network (VPN) connection. 
We developed a planification tool, named the SPeculoos Observatory sChedule maKer ({\fontfamily{pcr}\selectfont{SPOCK}}, which is optimised to the `long stare' observation strategy of SPECULOOS (see section \ref{sec:targets})). Its primary aim is to automatically generate and submit daily observing scripts (ACP observing plans) for the SSO, SNO, and SAINT-EX observatories. These are simple text files, which are linked to each other and allow to pre-define the nights observing, flat and calibration schedule. Its source code is available on Github\footnote{\url{https://github.com/educrot/SPOCK}}. These pre-defined observation plans allow us to automatically start the observations at any time in the night when weather conditions are fine. In case of unsafe weather, the ACP weather server allows to stop the observations and close the dome in order to avoid damage to the equipment. Additionally, the dome is directly connected to the weather station which offers a redundant way to automatically close the dome in such conditions.

To allow automatic continuation of observations, we have developed a script that monitors the ACP weather server after the observations have been terminated due to high winds, clouds or high humidity. It implements specific conditions that need to be met during a specific waiting time to apply lower wind, cloud, or humidity limits than we use for the termination of observations. It allows us automatically to determine whether the conditions have improved and if it's safe to continue observations, using the pre-defined plans. In case rain or snow has been detected, an automatic continuation will not be possible for safety reasons.
This automated script has been tested and adopted for the SSO and SNO observatories and is in regular use for these sites. In principle, it can be used for any observatory, which is controlled by ACP.
All telescopes require a minimum of on-site maintenance. The host observatories provide emergency help in case of technical difficulties as well as regular check-ups to ensure continuous robotic operations.  

\section{Targets and Observations}
\label{sec:targets}
\subsection{Target selection}
The SPECULOOS target list contains a homogeneous selected sample of close-by low-mass stars and UCDs\cite{Sebastian20}. The targets have been selected as low-mass dwarfs starting from the Gaia DR2 point source catalogue\cite{Gaia2016, Gaia2018} which has been cross-matched with the 2MASS point-source catalogue\cite{2MASS}. During this cross-match we enforced the agreement between the two catalogues not only in terms of position but also in terms of effective temperatures inferred from different photometric indicators. 
The final selection of the SPECULOOS target catalogue is based on each targets individual parameters and its match to the SPECULOOS science goals. These are (i) finding rocky Earth-like planets around close and bright targets, that are well-suited for atmospheric characterisation with JWST and (ii) to provide a census of the short-period planet population of UCDs. For the brighter targets in this catalogue, we leverage on the efficiency to detect small planets within TESS. That means, whether these targets are monitored with the SPECULOOS network depends on whether the photometric quality of the TESS data allows to detect small transit signals. Such signals will be followed-up by the SPECULOOS telescope network. The 40\,pc list of late-type targets and SPECULOOS target list are publicly available\cite{Sebastian20} together with a detailed description of the target selection and survey strategy. The target list is divided into three non-overlapping observing programmes:
\begin{itemize}
    \item Programme~1: includes 365 targets for which a transiting Earth-like planet (same size, mass, irradiation, and atmospheric composition than the Earth) will allow atmospheric characterisation by means of transit transmission spectroscopy with JWST. For the brighter and earlier targets from this list SPECULOOS telescopes follow up any transit signal within TESS. Each of the remaining targets will be observed for at least 200\,hr in order to effectively survey any possible period up to the mid-habitable zone\cite{kopparapu13} of these stars with an average integrated phase coverage of 80\%.  
    \item Programme~2: includes 171 targets with spectral type M5 and later, which are not in Programme~1, but bright enough to allow a detection of Earth-sized temperate (here: irradiation of 4\,$S_{\oplus}$) planets -- like TRAPPIST-1b -- with TESS. The SPECULOOS telescopes follow-up any transit signal detected within TESS.
    \item Programme~3: includes 1,121 targets with spectral type M6 and later and aims to explore the occurrence rate for temperate planets of UCDs within 40\,pc. In order to detect short-period up to temperate planets -- like TRAPPIST-1b -- each of these targets will be observed for at least 100\,hr. This allows us to effectively survey any possible period smaller than 6 days with an average integrated phase coverage of 80\%.
\end{itemize}

\begin{figure}[ht]
  \centering
   \includegraphics[width=0.7\textwidth]{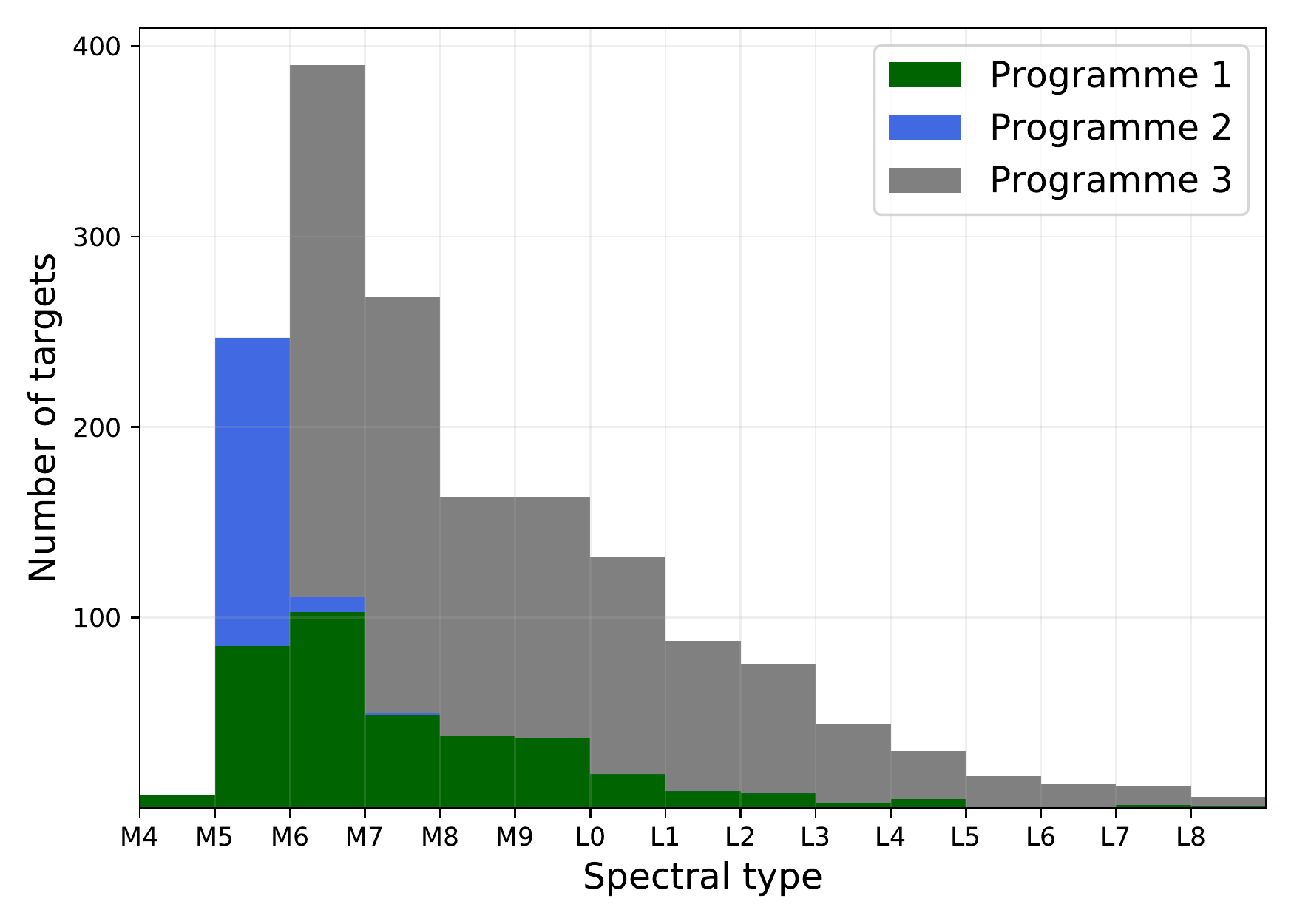}
   \caption{Spectral type distribution of all targets within the three SPECULOOS programmes.}   \label{fig:program}
\end{figure}

The full target catalogue comprises of 1,657 targets. As shown in Fig. \ref{fig:program}, the spectral type distribution of the SPECULOOS catalogue peaks at M6V with about 400 targets and decreases with later spectral types. The catalogue is incomplete for targets earlier than M6V because of the varying spectral type cut of the different programmes. Nevertheless, no cuts were introduced for later-type stars to maximise the catalogue completeness for UCDs.

\subsection{Scheduling}

Each night, we observe continuously one or two targets per telescope. Despite strategies with non-continuous observations are in use\cite{Nutzman2008,tamburo19}, this observation approach has been shown to be more reliable for the detection of short transit events (down to 15\,min for very-short-period $\leq$ 1\,d planets) among UCDs\cite{Delrez2018b,gibbs20}.

The SPECULOOS planification tool {\fontfamily{pcr}\selectfont  SPOCK} schedules targets completely automatically depending on their priority and observability. The priority is defined as the value of the S/N that has been used to select the targets for the different observing programs. The S/N for JWST transmission spectroscopy for an `Earth-like' planet in Programme~1, the TESS detection S/N for a temperate planet in Programme~2, and the SPECULOOS detection S/N for a temperate planet in Programme~3. The higher the corresponding S/N the higher is the targets priority. The priority of each target also depends on the completion factor. The completion ratio is defined as
\begin{equation*}
    r_{c}=\frac{hours_{observed}}{hours_{threshold}}, 
\end{equation*}
which embodies the fraction of hours of observation completed versus the number of hours required for each target. The required number of hours for each target is defined by the observing programme it belongs to. This completion ratio is used to  rank targets to favour the quick completion of ongoing targets, as opposed to starting new ones continually.

The observability is defined as the best visibility window for each target, based on its coordinates.
For each night, the selection process ranks all targets that are at their optimum visibility (respecting constraints imposed by the observatory such as Moon angular distance and altitude) and selects the one with highest priority. To implement those constraints, {\fontfamily{pcr}\selectfont  SPOCK} makes use of the {\fontfamily{pcr}\selectfont ASTROPLAN} package \cite{astroplan2018}, which is a flexible Python toolbox for astronomical observation planning and scheduling. It also optimises the annual period of the year for which the target is the most visible at a relatively low airmass, based on the completion ratio of the target and the optimisation of available observatories.

Targets that are observable all night long will be scheduled for the complete night. However, some targets have latitudes that do not allow for observation during all the available night time for the given site -- even at their peak of visibility -- such that observational gaps at the beginning or end of the night can appear for those targets. In that regard, an additional target is added to complement the schedule and avoid losing observing time. The observation durations of the two targets are set to be comparable, to prevent having overly short observation blocks (1\,hr or less). We say comparable because we do not exactly split the night in half; instead, we adopt a nightly observing duration adapted to each target's visibility (e.g. depending on Moon angular distance), which will shift from night to night during the observation period.
This situation of observing two targets per night is rather frequent and thus the most common observation mode for this survey.

One of the main roles of {\fontfamily{pcr}\selectfont  SPOCK} is to handle the coordination of multi-site observations. For instance, between two targets with similar priorities but one observable only from one site and the other from several sites, {\fontfamily{pcr}\selectfont  SPOCK} will choose the target that yields the most coverage. In addition, when possible, a one-hour overlap between observations from two different sites is scheduled to aid in the combination of the light curves. 

\section{Data reduction and quality}
\label{sec:pipeline}

\subsection{Data Reduction Pipelines}

The raw images from the SSO observatory are automatically transferred at the end of each night to the online ESO archive\footnote{\url{http://archive.eso.org/eso/eso_archive_main.html}}. These images are then downloaded to a server at the University of Cambridge (UK) the next day. In parallel, the raw images from the SNO observatory are directly transferred each day to the same Cambridge server.  

Images from the SSO and SNO observatories are then processed daily by the automatic SSO Pipeline\cite{murray20}. The SSO Pipeline is custom-built for the calibration and photometry requirements of the SPECULOOS survey, following a modular architecture similar to the Next-Generation Transit Survey (NGTS) pipeline\cite{wheatley2018}, and utilising the {\fontfamily{pcr}\selectfont  CASUTOOLS}\cite{Irwin2004} package of image processing tools (namely {\fontfamily{pcr}\selectfont  IMCORE, WCSFIT, IMSTACK} and {\fontfamily{pcr}\selectfont  IMCORE\_LIST}). The science images are reduced using standard methods of bias and dark subtraction and flat-field correction. We then use a local version of {\fontfamily{pcr}\selectfont  ASTROMETRY.NET} code \cite{lang2010} to cross-match our science images with reference catalogues built from the 2MASS catalogue to generate initial World Coordinate System (WCS) solutions. To refine these WCS solutions we perform source detection on our images and then, using the initial WCS solutions, cross-match these sources with the Gaia DR1 catalogue\cite{Gaia2016}, to further correct each image for translations, skews, scales, and rotations. For each field that is observed the pipeline requires an input catalogue, containing the positions of every source from which to extract aperture photometry data. We choose to have one, unique catalogue for every observed field to reference for all nights that field is observed. This allows us to monitor the photometry of all sources in this field consistently over long periods of time. To create this catalogue we perform source detection on a stacked image, generated from 50 images taken in the middle of the first night of observation of a field. We also cross-match this catalogue with Gaia DR2\cite{Gaia2018} to apply proper motion corrections on a night-by-night basis. There is also the facility to cross-match with other catalogues, such as 2MASS\cite{2MASS}. This catalogue then defines the central positions of apertures from which we measure raw photometry for each object in the field, for 13 different aperture radii. 

Once the raw, aperture photometry has been extracted, the SSO Pipeline corrects for ground-based systematics shared by other objects in the field by using an automated differential photometry algorithm. This iterative algorithm calculates an ‘artificial’ comparison light curve by weighting the sufficiently bright comparison stars according to their variability and distance to the target. Due to the survey's design, the redder comparison stars in the field are significantly fainter than the target, therefore we do not input any colour information into the weightings, to avoid introducing noise into the target's light curve. This mismatch in spectral type between the target and comparison stars leads to second-order differential extinction residuals imprinted on the target's light curve. To mitigate these residuals we implement a correction for precipitable water vapour from first principles, detailed in section \ref{pwv}.

SAINT-EX uses a different reduction process to SSO and SNO. The data are reduced by a custom pipeline, PRINCE (Photometric Reduction and In-depth Nightly Curve Exploration), that ingests the raw science and calibration frames and produces clean light curves. The PRINCE pipeline performs standard image reduction steps, applying bias, dark, and flat-field corrections. Astrometric calibration is conducted using Astrometry.net \cite{lang2010} to derive correct world coordinate system (WCS) information for each exposure. Photutils star detection \cite{bradley2019} is run on a median image of the whole exposure stack to create a pool of candidate stars in the field of view. Stars whose peak value in the largest aperture is above the background by a certain threshold, defined by an empirical factor times the median background noise of the night, are kept as reference stars for the differential photometric analysis. From the WCS information and the detected stars’ coordinates, the pipeline runs centroiding, aperture and annulus photometry on each detected star from the common pool, using LMFit \cite{newville2014} and Astropy \cite{astropy:2013,astropy:2018}, and repeats this for each exposure to obtain the measured light curves for a list of apertures. 

PRINCE then corrects for systematics in these light curves via two separate methods. The first is a simple differential photometry approach that corrects a star’s light curve by the median light curve of all stars in the pool except for the target star. The second method is a weighted PCA approach \cite{bailey2012}\footnote{\url{https://github.com/jakevdp/wpca}} with some added features. Outlying data points are removed by an iterative sigma-clipping procedure, where stars with a large fraction of the light curve as outliers are flagged. Stars which appear blended or have close neighbours are also flagged. In the PCA, data points of each star are weighted by the SNR, with flagged stars, the target and outliers having their weight set to zero. The PCA is then ran a second time, with uncertainties scaled such that a reduced $\chi^2$ for each star’s light curve equals unity. This has the effect of increasing the weight of well-behaved stars. 


\subsection{Photometric quality and the effect of precipitable water vapour}\label{pwv}
Observing in the NIR, necessary due to the faintness of very red targets in the optical, provides additional photometric challenges to those usually faced by ground-based observatories. For the SPECULOOS core program we use the \textit{I}+\textit{z} photometric filter for the majority of observations. This wavelength range is strongly affected by tellurics, most notably by atmospheric water absorption lines. As redder wavelengths are more readily absorbed by water than bluer wavelengths, the photometric impact will depend on the spectral energy distribution, and therefore increase for later spectral types. When we perform differential photometry we then introduce differential residuals due to the difference in spectral types between the red target and bluer comparison stars. The amount of water in the atmosphere can change rapidly and the resulting photometric residuals can be significant, of the order of $\sim$1\%. Mitigating this water vapour effect is therefore essential as it can mimic transit features in our light curves and complicate the long term variability studies of our targets.

Where they are available, we can use measurements of the precipitable water vapour (PWV) from the ground to directly probe the level of water absorption in the atmosphere, allowing us to model the Earth's atmospheric transmission. For the SNO observatory, 30\,minute cadence PWV measurements are already available from a GPS based system\cite{castro16}. It is planned to install and test a tau-monitor (Furuno) in 2021, which should provide more precise and accurate PWV measurements and with a higher cadence. For the SSO observatory we have access to high cadence zenith PWV measurements from the LHATPRO (Low Humidity And Temperature PROfiling radiometer) instrument at the Very Large Telescope in Cerro Paranal, situated approximately 150~m in altitude difference and 1.8~km in lateral distance from the SSO observatory \cite{kerber2012}. To model the atmospheric conditions at the SSO observatory we use the Sky-Calc Sky Model Calculator, a tool developed by ESO which is based on the Cerro Paranal Advanced Sky Model\cite{jones2013, noll2012}. This tool uses PWV and airmass values to provide a corresponding atmospheric transmission model. By taking into account the instrument response through the \textit{I}+\textit{z} filter, and using synthetic stellar spectra from PHOENIX\cite{husser2013}, we can predict the photometric impact for objects of different spectral types, for a given PWV value and airmass. We then correct the differential light curves by dividing by the predicted PWV effect. This process\cite{pedersen21} is implemented as part of the SSO Pipeline differential photometry algorithm. 

\begin{figure}[ht]
  \centering
   \includegraphics[width=0.7\textwidth]{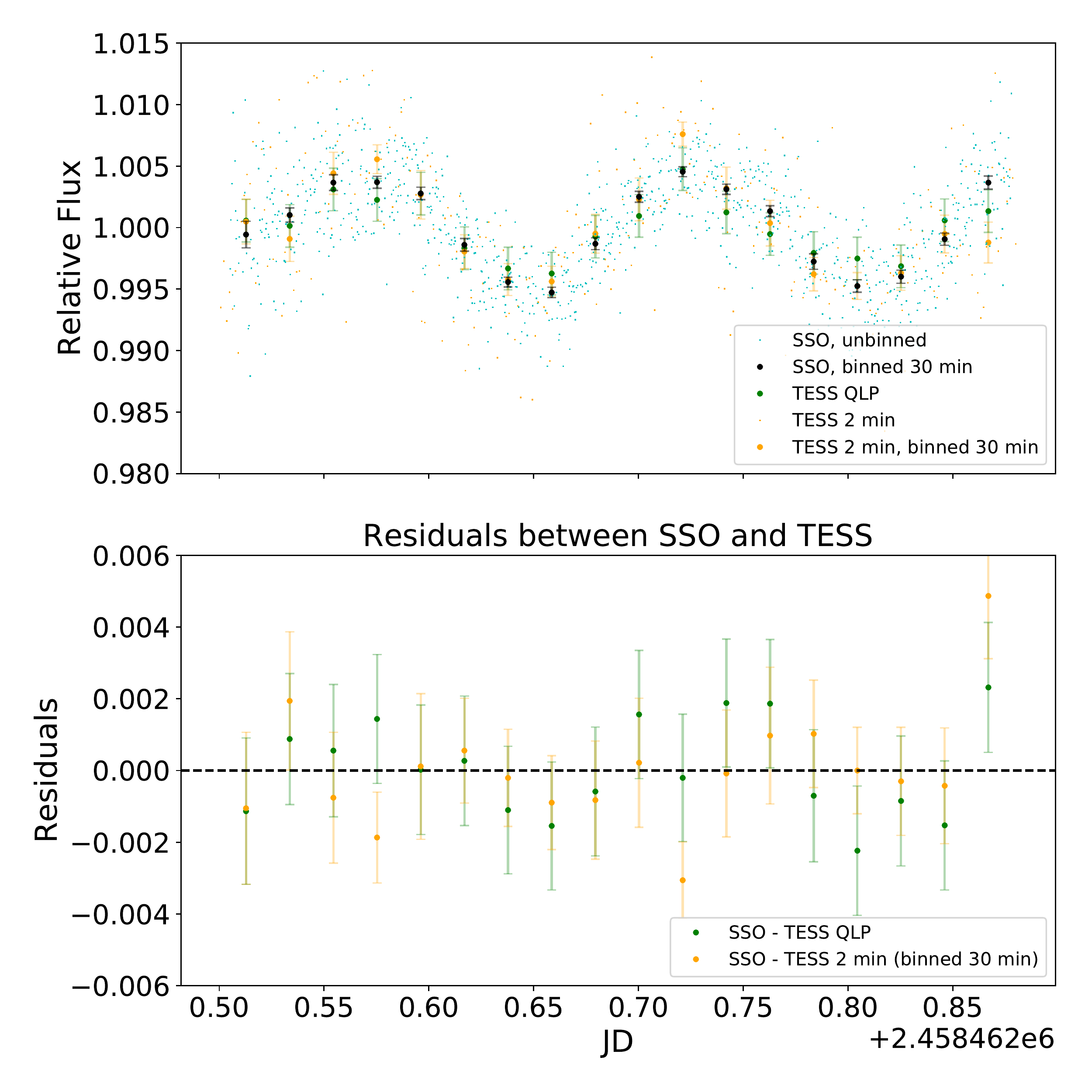}
   \caption{Top: Comparison of light curves from the SSO observatory and TESS for an M6V object (J = 10.3mag) on 2018 December 10. The differential light curve from a single SSO telescope is in cyan, with 30\,minute binned data in black, the light curve from TESS 2\,minute cadence data is in orange and MIT Quick Look Pipeline 30\,minute cadence data is in green. Bottom: The residuals between the SSO light curve and the two TESS light curves.}   \label{fig:tess}
\end{figure}

To assess the photometric quality of the SSO observatory, we measured the fractional RMS of our target light curves for 30\,minute bins, from January 1st (start of official scientific operations) to September 18th, 2019\cite{murray20}. This data sample contained 98 targets observed over 179 nights of observation with the SSO telescopes. We performed no detrending, correction for intrinsic variability or removal of sub-optimal observing conditions. In this analysis, we demonstrated that we are reaching sub-millimag precisions for $\sim$30\% of light curves (with a median precision of 1.5\,mmag), and up to 0.26\,mmag for the brightest objects. There have been recent developments to the SSO Pipeline to clean the final light curves, including flagging frames where the observing conditions significantly affect the photometry. This process also includes ongoing work to identify and remove flares\cite{murray21}, and modelling of the stellar variability of our targets from long term photometric monitoring. Removing the activity of our targets will allow us to further refine the photometric quality of our facilities and improve our detection efficiency for small planets.

In addition, we compared simultaneous observations of a variable M6V object with a single SSO telescope and TESS 2\,minute and 30\,minute cadence observations (Figure \ref{fig:tess})\cite{murray20}. There was excellent agreement between the three datasets, with the SSO light curve exhibiting the least white noise. As TESS is not optimised for UCDs, we expect that the quality of the SSO light curves will exceed TESS when we observe cooler and redder objects, however for the bright M5V and M6V objects the quality of the photometry will be comparable. This demonstrates how we can utilise the synergy between SPECULOOS and TESS to optimise the detection of Earth-sized planets, as in Programme~2. This work also highlights the impressive photometric performance of the SSO observatory, especially for quiet targets observed on nights with good observing conditions.

\section{SPECULOOS portal}
\label{sec:portal}

To enable easy access to the SSO Pipeline’s output, a web-based service and interface was designed, called PORTAL (Pipeline Output inteRacTion Analysis Layer).  It was built using a common backend stack – LAMP (Linux, Apache, MySQL, PHP/Python), with a Vue.js, Plotly, and D3.js based frontend. Its main operation permits quick and interactive visualisation of nightly light curves from the SSO and SNO observatories. It also serves as a RESTful API, which allows members of the SPECULOOS consortium to download and analyse any of the pipeline’s outputs and connected metadata.

\subsection{Backend}

The backend bridges the link between SSO Pipeline data products and the user-friendly frontend interface, by behaving as a RESTful API. The pipeline’s output directories are navigated by the backend, using Gaia DR2 source IDs as the main search parameter, with the option to further refine by specifying date, telescope, and filter – with arrays of values and wildcards permitted in a search query.  To retrieve nightly observation data, the backend processes multiple text files (one per aperture) containing nightly differential light curves (PWV and non-PWV corrected) and metadata (relative RA and DEC movement, FWHM, PSF, sky level, and airmass), and formats them into a single JSON structure.

For larger queries, a SQL database was implemented. It contains all the nightly observation data, as well as environment/telescope specific metadata. The tables are updated whenever a directory is updated by the pipeline.  The database also stores SPECULOOS’ observation history (based on the pipeline’s output directories) and the observation schedule (provided by {\fontfamily{pcr}\selectfont  SPOCK}’s output files). The SQL database also stores user submitted flags and comments on nightly data provided by the frontend. 

The total number of hours a target has been observed for, as processed by the pipeline, can be also queried. The backend also produces low resolution videos from the raw 2K~$\times$~2K images acquired during an observation, which forms part of a target’s detailed observation view produced by the frontend.

\subsection{Frontend}

\begin{figure}[H]
  \centering
   \includegraphics[width=0.7\textwidth]{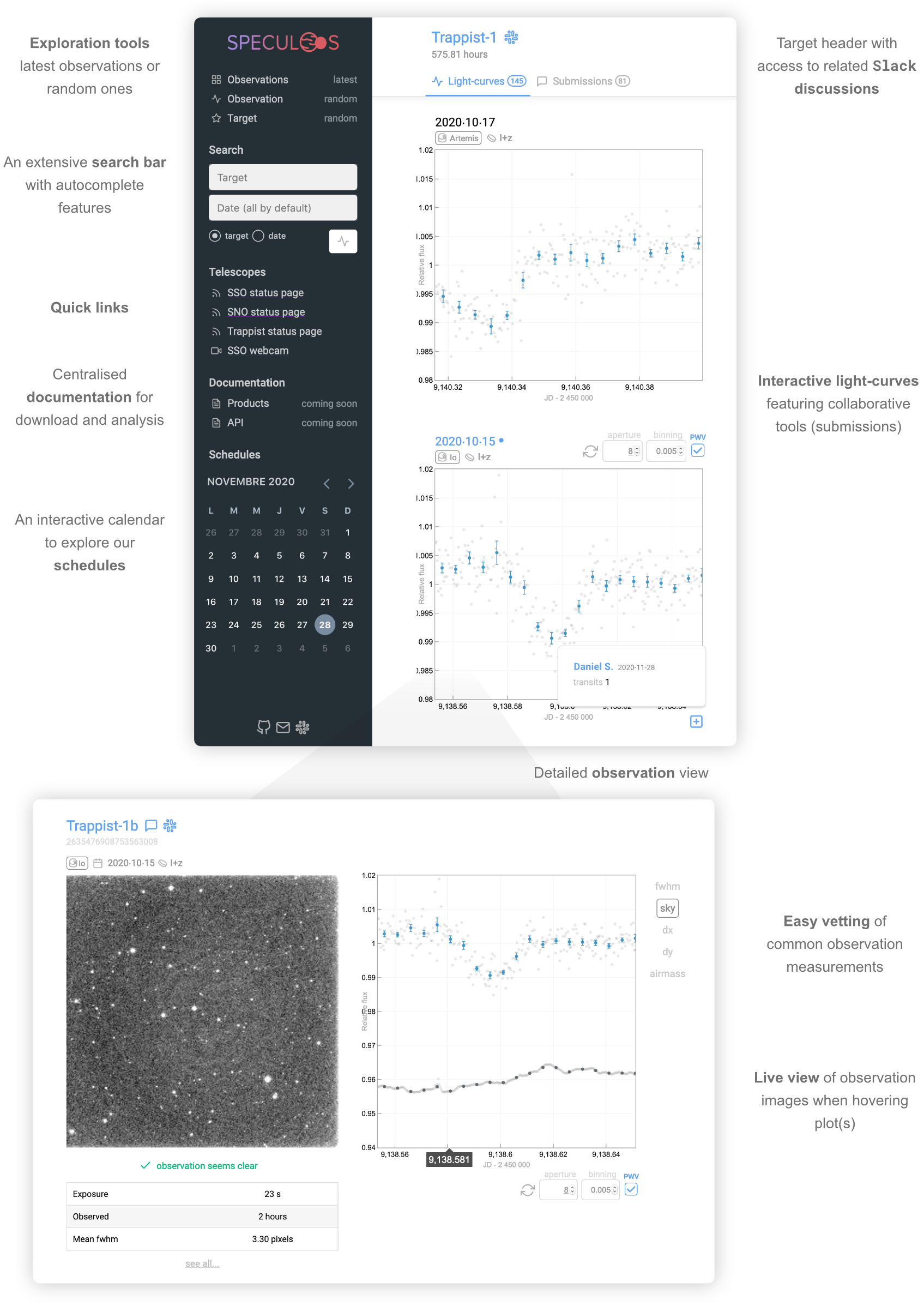}
   \caption{A descriptive view of PORTAL's web interface. Top: A target specific view is shown for Trappist-1, with an example set of interactive light curves and a user submitted flag. Bottom: A detailed observation view for a particular night for Trappist-1b. The collection of tools PORTAL offers allows for an intuitive and collaborative exploration of differential photometry data, which can be downloaded for extensive offline analysis via the backend's RESTful API.}   \label{fig:portal}
\end{figure}

The frontend is a user-centred designed interface for displaying differential photometry data. It allows navigation and interaction with observations made by all of SPECULOOS’ facilities, on a target and nightly basis.  The user-friendly interface, as shown in Figure~\ref{fig:portal}, has a vertical menu panel and an interchanging main pane.

The navigation panel contains a search function with autocomplete capabilities. A user can search with either a target’s Gaia DR2 source ID or the shorter SPECULOOS target ID. By default, the search function will return all the light curves for a particular target, ordered by newest date first. Nightly light curves are presented within interactive scatter plots, with raw and binned data. In its default view, the scatter plots are displayed with PWV corrected light curves (if available) using an aperture value pre-determined by the pipeline, within the differential flux range of [0.98,1.02], and a binned period of 0.005~JD. A user can quickly interact with the data by zooming in/out and toggling between apertures, binned periods, and the PWV applied correction. If a user spots an interesting feature, such as a flare, a transit feature, a type of variability, or an issue with the data, they can tag the data, as well as submit a comment to be connected with the observation. On hover, one can view previous submissions, or one can view all submissions for a target on a separate tab.

To vet the data further, one can access a target’s detailed observation view for a night, as exemplified in the bottom of Figure~\ref{fig:portal}. Here, one can explore metadata, such as relative RA and DEC movement, FWHM, sky level, and airmass. Some of the extended metadata are quality checked for anomalies, which are used to inform the user.  A video of the observation is also present, which allows one to navigate images of the night, triggered by hovering over different parts of the light curve. Lastly, one can automatically open Slack channels to allow for further discussions of the target.

To promote further vetting of data, a user can view the latest light curves, a random target’s light curves, or a detailed observation view for a random target and night. The schedule can also be navigated by date, and if the target has been observed, it links to its respective detailed observation view. The final links on the navigation panel are to the telescopes’ status pages, and a timelapse view of SSO's widefield night-sensitive webcam.

In future, we will continue to improve the usability and ease of the platform. One planned improvement is to allow users to view the comparison stars’ light curves used as part of the pipeline, and their position on a stacked image of the night.  We are also planning to integrate analysis tools that are currently being developed by consortium members to further vet nightly observations.

\section{Moving target pipeline for SPECULOOS}
\label{sec:move}

The main goal of this package under development is to implement a suit of programs optimised to find solar system moving objects (Main Belt asteroids, Near Earth asteroids, comets, Trans-Neptunian Objects objects, etc.), extract their astrometric positions and magnitude, report them to the MPC (Minor planet Center) and build their light curves. This pipeline will be included in the dataflow to automatically search the data when the night is over. It is based on a new software package, the SSOS tool\cite{mahlke19}, complemented by PHOTOMETRYPIPELINE\cite{mommert17} that have been adapted and optimised for the TRAPPIST data\cite{jehin11} and now the SPECULOOS project.\\
The development and implementation of this independant pipeline is driven by a set of scientific goals. The main goal is to retrieve the light curves of hundreds of known asteroids that are serendipitously recorded in the course of the SPECULOOS monitoring. Indeed a lot of Main Belt asteroids are expected to be present in the archival images down to magnitude 20 and the many nights spend continuously on each field allow to obtain long photometric series for slow moving objects. These light curves can be used to deduce the rotation periods of the asteroids because of their irregular shapes, most of them having periods between 2.5 and 20\,hr. Categories of special interest are the slow rotators (period $> 100$\,hr), a population that is poorly studied and might be biased because of lack of long photometric studies\cite{marciniak15}. These light curves can also reveal very fast rotators (period $< 2.4$\,hr) thanks to the fast cadence of observations of typically 60 seconds. Such rare asteroids are also very interesting to set up the limit between the rubble pile and single rocky body limit for asteroids\cite{monteiro20}. This sample should hopefully reveal also a bunch of new binaries as they represent about 15\% of the asteroids population\cite{margot15}. An other use of such data, in case of light curves taken at different phase angles or to complement missing phase angle for specific objects, allow to derive their precise shape after 3D reconstruction\cite{kaasalainen02}. The precise astrometric 
measurements of all the objects detected by the pipeline will be submitted to the Minor Planet Center (MPC) and will be used to compute more precise orbits.

Finally, the pipeline could lead to the discovery of new asteroids, and maybe if we are lucky, as they are rare, new comets, large Trans-Neptunian Objects or even interstellar objects, recorded in the images by chance. There are dedicated surveys that are doing such searches every clear nights, but still a serendipitous discovery is always possible as the telescopes just need to point at the right place at the right time. As SPECULOOS telescopes are looking at any location in the sky, and not specifically the ecliptic, it is possible to find interesting objects on high inclination orbits that have escaped detection until today.

\section{Results and Discussion}
\label{sec:discussion}

\subsection{SPECULOOS core programme}
Observations of the SPECULOOS core programme are performed with the SPECULOOS network every suitable night. Since the start of observations 265 targets from the target list have been observed with 30\% of them more than 100h photometric data have been collected. Most of the observed targets are programme 1 targets, thus bright and close-by targets. For 5\% of the programme 1 targets, the observations have been completed, hence for any possible planet within or closer than the habitable zone on average 80\% of the orbital phase has been covered.
The actual completion factor is higher, since high S/N TESS light curves of the brightest targets are analysed in parallel. Given this synergy, we expect to complete our most time intensive programme 1 in less than four years. Using current models for planetary systems of UCDs, we expect\cite{Sebastian20} to find about a few dozens temperate, rocky planets within the SPECULOOS survey, with a handful of them being amenable for atmospheric characterisation with JWST. 

The SPECULOOS database delivers high quality light curves for UCDs within 40pc, making it possible to study their photometric variability, rotation periods and flare activity. The SPECULOOS PORTAL offers the SPECULOOS science team optimised tools to visually inspect light curves on a daily basis, including a community driven approach to motivate reporting of interesting features.
Furthermore, the moving target pipeline will make use of the SPECULOOS field of view to analyse hundreds of known asteroids and possibly identify new, unknown solar system objects.

\subsection{Annex programmes}

Beside conducting observations of targets from the SPECULOOS input catalog, a fraction of the available observing time of the SPECULOOS network is used to carry out different science goals, also so-called annex programmes. These are not part of the SPECULOOS core science case, and have revealed a wealth of different outcomes since the commissioning of our telescopes.
A large annex programme is the support of space based transit search surveys such as K2 and TESS through the follow-up of transit candidates of late-type dwarfs.
Following-up photometric data from K2, using SSO and SNO observatories revealed the detection of $\pi$ Earth\cite{Niraula2020}, a transiting Earth-sized ($0.95\,R_{\oplus}$) planet around the mid-M dwarf K2-315b in 57\,pc distance, which is well suited for comparative terrestrial exoplanetology. The first discovery from the SAINT-EX observatory has been made possible by following-up the TESS transit candidate TOI-1266\cite{Demory2020}, which revealed a system that hosts a super-Earth and a sub-Neptune around a M3 dwarf. The outer planet TOI-1266c has an irradiation level similar to that of Venus and is thus a favourable target for future atmospheric characterisation. SSO photometry was involved in the discovery of a system of three close-in terrestrial planets, that have been found orbiting the M3 dwarf L 98-59\cite{kostov2019}. Due to the relative brightness of the host star, these planets are benchmark targets for future atmospheric characterisation of terrestrial planets.

The SPECULOOS network has also been involved in projects quite different from planet transit characterisation, e.g. studying the properties of low-mass stars and brown dwarfs. Notably, an eclipsing binary brown dwarf with a tertiary brown dwarf companion 2MASSW J1510478-281817\cite{triaud20} was detected using the SSO photometry. It is one of only two double-lined, eclipsing brown dwarf binaries known today and, thus, provides a rare benchmark to constrain the masses, radii, and ages of such objects.

SPECULOOS photometry is playing an important part in following up and putting constraints on the physical origin of a new class of rapid rotating low-mass stars showing chromatic, complex modulated light curves\cite{Guenther2020} or to characterise low-mass eclipsing binaries within the EBLM project\cite{Boetticher2019}.



\acknowledgments 
 
The research leading to these results has received funding from the European Research Council (ERC) under the European Union's Seventh Framework Programme (FP/2007--2013) ERC Grant Agreement n$^{\circ}$ 336480, from the ARC grant for Concerted Research Actions financed by the Wallonia-Brussels Federation, from the Balzan Prize Foundation, and from F.R.S-FNRS (Research Project ID T010920F). MG is F.R.S.-FNRS Senior Research Associate.
This research is also supported work funded from the European Research Council (ERC) the European Union’s Horizon 2020 research and innovation programme (grant agreement n$^{\circ}$ 803193/BEBOP), from the MERAC foundation, and through STFC grants n${^\circ}$ ST/S00193X/1 and ST/S00305/1. This work was also partially supported by a grant from the Simons Foundtion (PI Queloz, grant number 327127). J.d.W. and MIT gratefully acknowledge financial support from the Heising-Simons Foundation, Dr. and Mrs. Colin Masson and Dr. Peter A. Gilman for Artemis, the first telescope of the SPECULOOS network situated in Tenerife, Spain. B.V.R. thanks the Heising-Simons Foundation for Support. MNG acknowledges support from MIT's Kavli Institute as a Juan Carlos Torres Fellow.
This work has made use of data from the European Space Agency (ESA) mission
{\it Gaia} (\url{https://www.cosmos.esa.int/gaia}), processed by the {\it Gaia}
Data Processing and Analysis Consortium (DPAC,
\url{https://www.cosmos.esa.int/web/gaia/dpac/consortium}). Funding for the DPAC
has been provided by national institutions, in particular the institutions
participating in the {\it Gaia} Multilateral Agreement.
This publication makes use of data products from the Two Micron All Sky Survey,
which is a joint project of the University of Massachusetts and the Infrared Processing 
and Analysis Center/California Institute of Technology, funded by the National Aeronautics 
and Space Administration and the National Science Foundation.
This research has made use of the SIMBAD database,
operated at CDS, Strasbourg, France, This research has made use of the VizieR catalogue access tool, CDS, Strasbourg, France, This research made use of Astropy,\footnote{http://www.astropy.org} a community-developed core Python package for Astronomy \cite{astropy:2013, astropy:2018}.

\bibliographystyle{naturemag}

\end{document}